\documentclass[conference]{IEEEtran}
\usepackage{amsmath}
\usepackage{amsfonts}
\usepackage{cite}
\usepackage{amssymb}
\usepackage{graphicx}
\usepackage[usenames,dvipsnames]{color}
\usepackage{epstopdf}
\usepackage[all]{xy}
\usepackage[]{mcode}

\begin{document}
\title{Outage Performance Analysis of Multicarrier Relay Selection for Cooperative Networks}
\author{\IEEEauthorblockN{Shuping Dang, Justin P. Coon, Gaojie Chen and David E. Simmons}
\IEEEauthorblockA{Department of Engineering Science, University of Oxford, Oxford, UK, OX1 3PJ\\
Email: \{shuping.dang, justin.coon, gaojie.chen, david.simmons\}@eng.ox.ac.uk}\\ \ \\}

\maketitle

\begin{abstract}
In this paper, we analyze the outage performance of two multicarrier relay selection schemes, i.e. bulk and per-subcarrier selections, for two-hop orthogonal frequency-division multiplexing (OFDM) systems.  To provide a comprehensive analysis, three forwarding protocols: decode-and-forward (DF), fixed-gain (FG) amplify-and-forward (AF) and variable-gain (VG) AF relay systems are considered. We obtain closed-form approximations for the outage probability and closed-form expressions for the asymptotic outage probability in the high signal-to-noise ratio (SNR) region for all cases. Our analysis is verified by Monte Carlo simulations, and provides an analytical framework for multicarrier systems with relay selection.
\end{abstract}


\IEEEpeerreviewmaketitle

\section{Introduction}
Cooperative communications and relay-assisted systems have attracted a large amount of attention since they were proposed and comprehensively analyzed in \cite{1362898}. In such a network, relays are employed as intermediate communication nodes to assist the transmission between source and destination, so that the communications can be maintained when the direct transmission link between source and destination undergoes deep fading \cite{7060199,5733966}. It has been proved that with the help of relays and proper forwarding protocols, the network coverage expansion, energy efficiency, system reliability and quality of service (QoS) can be significantly enhanced \cite{5196673, 6612901, 4027619,5765728}. In particular, relay selection can further enhance the system performance and obtain an extra diversity \cite{4657317}. A variety of selection schemes have been proposed and analyzed for different types of relays and channel conditions \cite{5756421,6329298, 5555992,wang2015outage,6678968, 6220222}. 

Meanwhile, multicarrier communications, especially orthogonal frequency-division multiplexing (OFDM), is also proposed and utilized in relay-assisted networks, which is capable of providing a better system performance and bandwidth efficiency over frequency selective channels \cite{841722,4086395}.  But relay selection in multicarrier systems is not straightforward, because this is a two-tier selection/allocation problem involving relay selection and subcarrier allocation. In \cite{4489212}, two relay selection schemes for OFDM systems, i.e. bulk selection (a single relay is selected for transmission on all subcarriers) and per-subcarrier selection (selection is treated independently for each subcarrier, thus, potentially, leading to transmission via multiple relays) are proposed and analyzed. 

However, all previous works on multicarrier relay selection are carried out based on an assumption that the outage condition regarding each subcarrier is treated individually, which corresponds to the block fading channel model when a user is allocated a contiguous block of carriers that lie within a coherence bandwidth. To be more realistic, we should consider the parallel fading channel model that general OFDM systems are well approximated by. The parallel fading channel model is utilized to model a fading channel consisting of a finite number of flat independent and identically distributed (i.i.d.) fading subchannels \cite{6353210}. The most unique property of the parallel fading channel relative to the conventional block fading channel is the definition of outage probability. Considering the coding over the parallel fading channel, the outage probability is defined as the probability that the mutual information of all subchannels is smaller than a target transmission rate \cite{tse2005fundamentals}. However, the mathematical tractability of the outage probability over the parallel fading channel is rather poor, because the distribution of the summation of a finite number of random variables cannot be derived in a generic closed-form expression \cite{6111188}. In \cite{293655}, an oversimplified case of a parallel fading channel with only two subchannels is analyzed and an integral formula for the outage probability is given. Also, upper and lower bounds as well as an approximation for the outage probability for any number of subchannels are determined in \cite{6353210} and \cite{7312903}, respectively. 

On the other hand, to the best of the authors' knowledge, the work considering multicarrier relay selection for two-hop OFDM systems approximated by parallel fading channel model has not been fully treated. To fill this gap, we analyze this scenario in a Rayleigh fading condition. Note, the method applied in this paper can be easily tailored to analyze other channel conditions.

The main contributions of this paper are summarized infra:
\begin{itemize}
\item We obtain closed-form generic approximations for outage probabilities of two-hop OFDM systems when applying bulk and per-subcarrier relay selection schemes.
\item We derive closed-form approximations for the outage probability of decode-and-forward (DF), fixed-gain (FG) amplify-and-forward (AF) and variable-gain (VG) AF relay systems with bulk and per-subcarrier relay selection schemes.
\item We obtain closed-form asymptotic expressions for the outage probability with different selection schemes at high SNR and also derive diversity gains.
\end{itemize}

The rest of this paper is organized as follows. In Section \ref{sm}, the system model and fundamentals are given. Then, the outage performance is analyzed in Section \ref{opa}. Specific applications including DF, FG AF and VG AF relay systems are investigated in Section \ref{app}. The analysis is numerically verified by Monte Carlo simulations and further discussed in Section \ref{nr}. Finally, Section \ref{c} concludes the paper.

\section{System Model}\label{sm}

\subsection{Parallel fading channel and outage probability}
In this paper, we consider a two-hop OFDM system with $K$ orthogonal subcarriers and $M$ relays gathered in a relay cluster, the size of which is relatively small compared to the distance between source and destination. Therefore, $K$ parallel fading subchannels are constructed at each hop, and for each subcarrier, there are $M$ interfering channels. We denote the sets of subcarriers and relays as $\mathcal{K}=\{1,2,\dots,K\}$ and $\mathcal{M}=\{1,2,\dots,M\}$. Also, we assume the entire two-hop OFDM system operates in a half-duplex protocol and the direct transmission link between source and destination does not exist due to deep fading. As a result, two orthogonal time slots are required for one complete transmission from source to destination via relay(s).  Here, we denote i.i.d. Rayleigh channel coefficients in the first and second hops by $h_1(m,k)$ and $h_2(m,k)$ $\forall m\in\mathcal{M}$ and $k\in\mathcal{K}$. The channel gain $|h_i(m,k)|^2$, where $i\in\{1,2\}$, is exponentially distributed with the mean of $\mu_i$. Therefore, its probability density function (PDF) and cumulative distribution function (CDF) are
\begin{equation}\label{pdfchannelsingle}
f_{|h_i|^2}(x)={e^{-x/\mu_i}}/{\mu_i}~\Leftrightarrow~F_{|h_i|^2}(x)=1-e^{-x/\mu_i}.
\end{equation}

Subsequently, the end-to-end SNR transmitted on the $k$th subcarrier and forwarded by the $m$th relay is denoted as $\gamma(m,k)$. Accordingly, the outage probability over the parallel fading channel can be defined as 
\begin{equation}\label{defoutage}
\begin{split}
P_{out}(s)=\mathbb{P}\left\lbrace \sum_{k=1}^{K} \ln (1+\gamma(m_k,k))<s \right\rbrace
\end{split}
\end{equation}
where $\mathbb{P}\left\lbrace\cdot \right\rbrace$ denotes the probability of the enclosed; $m_k$ is the index of the relay selected to forward the $k$th subcarrier; we also let $s=2\xi$ for the convenience of the following analysis, where $\xi$ is the mutual information outage threshold\footnote{The factor `2' comes from the fact that two time slots are required for one complete transmission in two-hop systems.}. In addition, we assume all noise statistics are assumed to be zero-mean, complex Gaussian random variables with variance $N_0/2$ per dimension, from which the noise power can be expressed by $N_0$.

\subsection{Forwarding protocols}
We assume equal bit and power allocation schemes are applied and the average transmit power per subcarrier at source and relay is the same, denoted by $P_t$.  Therefore, the instantaneous end-to-end SNR of the $k$th subcarrier forwarded by the $m$th relay using a DF protocol is written as
\begin{equation}\label{snrdecodeandforward}
\gamma_{DF}(m,k)={\bar{\gamma}\min\left(|h_1(m,k)|^2,|h_2(m,k)|^2\right)},
\end{equation}
where $\bar{\gamma}=P_t/N_0$. Similarly, for FG AF relaying that is blind to all channel conditions and is only able to amplify the received signal by a fixed gain, the instantaneous end-to-end SNR is written as 
\begin{equation}\label{snrfixedgain}
\gamma_{FG}(m,k)=\frac{\bar{\gamma}^2|h_1(m,k)|^2|h_2(m,k)|^2}{\bar{\gamma}\mu_1 +\bar{\gamma}|h_2(m,k)|^2+1}.
\end{equation}
For VG AF relaying which is able to estimate channel conditions and amplify accordingly, the instantaneous end-to-end SNR is thereby
\begin{equation}\label{snrvariablegain}
\gamma_{VG}(m,k)=\frac{\bar{\gamma}^2|h_1(m,k)|^2|h_2(m,k)|^2}{\bar{\gamma}|h_1(m,k)|^2+\bar{\gamma}|h_2(m,k)|^2P_t+1}.
\end{equation}

\subsection{Relay selection schemes}

\subsubsection{Bulk selection}
By the bulk selection scheme, the source only selects \textit{one} out of $M$ relays by which all subcarriers are forwarded according to the selection criterion 
\begin{equation}\label{bsc}
\mathcal{L}^{bulk}=\arg\max_{m\in\mathcal{M}}\left\lbrace\sum_{k=1}^{K} \ln (1+\gamma(m,k))\right\rbrace,
\end{equation}
where $\mathcal{L}^{bulk}$ is the set (of cardinality one) denoting the only selected relay. This selection scheme is easy to implement and will not involve an overcomplicated coordination protocol, because there is only one selected relay. However, it is obvious that the outage performance cannot be optimized in this case.

\subsubsection{Per-subcarrier selection}
\begin{figure}[!t]
\centering
\includegraphics[width=3.5in]{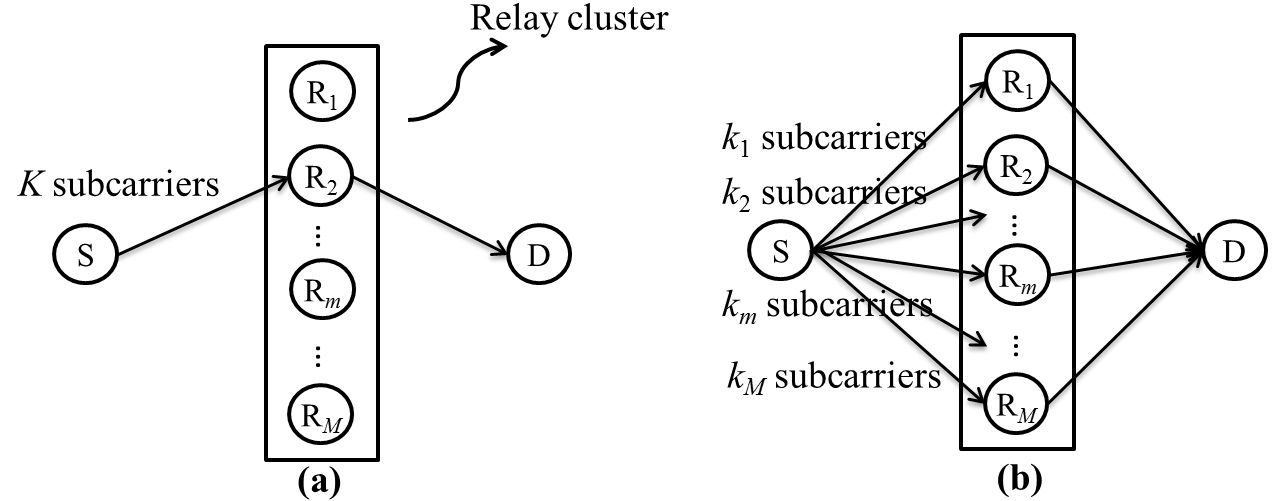}
\caption{Illustration of (a) bulk, (b) per-subcarrier relay selection schemes for single source, single destination and $M$ clustered relays with $K$ subcarriers. Note, $K=\sum_{m=1}^{M}k_m$ for $k_m\in \mathbb{N}$, $\forall ~1\leq m\leq M$.}
\label{sys}
\end{figure}

The per-subcarrier selection scheme selects $L$ relays from $M$ relays in a per-subcarrier manner and $1 \le L\le \min(M,K)$. Therefore, the relay selected to forward the $k$th subcarrier is individually determined by 
\begin{equation}\label{ptsc}
\begin{split}
l^{ps}(k)=\arg\max_{m\in\mathcal{M}}\ln (1+\gamma(m,k))
=\arg\max_{m\in\mathcal{M}}\gamma(m,k)
\end{split},
\end{equation}
where $l^{ps}(k)$ is the set of the selected relay corresponding to the $k$th subcarrier. Then, this selection process will be repeatedly applied for all subcarriers and finally $L$ relays are selected. The set of all $L$ selected relays is denoted by $\mathcal{L}^{ps}=\bigcup_{k=1}^{K}\left\lbrace l^{ps}(k) \right\rbrace$. Note, it is allowed that $l^{ps}(k)=l^{ps}(n)$ for $k\neq n$, i.e. one relay is allowed to assist in forwarding two or more subcarriers at the same time. Obviously, this selection scheme is optimal in terms of outage performance, but this optimality will result in a high system complexity \cite{6356932}. For illustration purposes, these two selection schemes are illustrated in Fig. \ref{sys}.

\section{Outage Performance Analysis}\label{opa}
\subsection{Bulk selection}
According to (\ref{defoutage}) and (\ref{bsc}), the \textit{a posteriori} outage probability with bulk selection can be defined by
\begin{equation}\label{21dsa1d2311bbbbs}
\begin{split}
&P_{out}(s)=\mathbb{P}\left\lbrace  \max_{m\in\mathcal{M}}\left\lbrace\sum_{k=1}^{K} \ln (1+\gamma(m,k))\right\rbrace<s \right\rbrace\\
&=\prod_{m=1}^{M} \mathbb{P}\left\lbrace \sum_{k=1}^{K} \ln (1+\gamma(m,k))<s\right\rbrace\overset{(a)}{=}(F_I(s))^M,
\end{split}
\end{equation}
where $F_I(s):= \mathbb{P}\left\lbrace  \sum_{k=1}^{K} \ln (1+\gamma(m,k))<s\right\rbrace$, $\forall m\in \mathcal{M}$ and $\forall k\in \mathcal{K}$; (a) is valid, because all parallel subchannels are assumed to be i.i.d.

Assume the CDF and PDF of $\gamma(m,k)$ are $F_{\gamma}(s)$ and $f_{\gamma}(s)$, respectively. Now following the \textit{Proposition 1} proved in \cite{7312903}, we also make a hypothesis that $f_{\gamma}(s)$ can be expanded as a power series about zero as
\begin{equation}\label{56465122211}
f_{\gamma}(s)=s^q(g_0+g_1s+O(s^2)),
\end{equation}
where $q$ is a non-negative integer representing the inherent subchannel diversity; $g_0$, $g_1$ are non-zero functions of $\bar{\gamma}$ and satisfy the condition $g_1(\bar{\gamma})=o(g_0(\bar{\gamma}))$ when $\bar{\gamma}\rightarrow \infty$. This is a common assumption applicable to most two-hop channel cases \cite{7445895}. Then, $F_I(s)$ can be written as \cite{7312903}
\begin{equation}
\begin{split}
&F_I(s)=\frac{S_0(K,s,q)^{Kq+1}}{(S_1(K,s,q)-S_0(K,s,q))^{Kq}}\\
&\times f_\gamma\left(\frac{S_1(K,s,q)-S_0(K,s,q)}{S_0(K,s,q)}\right)^K+O\left(\frac{1}{\bar{\gamma}^{K(q+1)+2}}\right).
\end{split}
\end{equation}

The coefficient $S_{x}(K,s,q)$ is defined as
\begin{equation}\label{dasj5645lopqs}
\begin{split}
&S_x(K,s,q):= \sum_{j=0}^{q}\binom{q}{j}(-1)^j\sum_{\substack{a_p\in\mathbb{N},~\forall 0 \leq p \leq q \\ \sum_{p=0}^{q}a_p=K-1}}\binom{K-1}{a_0,a_1,\dots,a_q}\\
&\times \left[\prod_{p=0}^{q} \left(\binom{q}{p}(-1)^p\right)^{a_p} \right]\cdot\left[\frac{1}{2\pi i}\int_{C}\frac{e^{sz}\mathrm{d}z}{\prod_{k=0}^{K}(z-\beta_k)}\right],
\end{split}
\end{equation}
where
$\beta_k$ is the $k$th element in
\begin{equation}
\mathbf{B}=(0,\underbrace{q+1,\dots,q+1}_{a_0 \mathrm{terms}},\underbrace{q,\dots,q}_{a_1 \mathrm{terms}},\dots,\underbrace{1,\dots,1}_{a_q \mathrm{terms}},q+1+x-j)
\end{equation}
and $C$ is the contour which encloses all poles of the integrand.

It should be noted that $\frac{1}{2\pi i}\int_{C}\frac{e^{sz}\mathrm{d}z}{\prod_{k=0}^{K}(z-\beta_l)}$ is equivalent to the inverse Laplace transform of ${1}/{\prod_{k=0}^{K}(z-\beta_k)}$, which can be expressed by the closed form as follows \cite{erdelyi1954tables}:
\begin{equation}
\begin{split}
&\frac{1}{2\pi i}\int_{C}\frac{e^{sz}\mathrm{d}z}{\prod_{k=0}^{K}(z-\beta_k)}=\sum_{k=0}^{K}\left[\frac{e^{\beta_k s}}{\prod_{\substack{1\leq n\leq K\\n\neq k}}(\beta_k-\beta_n)}\right].
\end{split}
\end{equation}
However, this closed-form expression given above is only valid if all poles of the integrand on the are simple (first-order poles), i.e. $\beta_n \neq \beta_k$, for $n\neq k$. Closed-form expressions exist for the cases with higher-order poles, but cannot be written as a general form. This is the reason why we still keep the integral form in (\ref{dasj5645lopqs}) for generality. Another reason for keeping the integral form is because this integral form can be efficiently evaluated by computer-based simulations using the residue theorem \cite{davies2012integral}.

As a result, the \textit{a posteriori} outage probability by bulk selection as defined in (\ref{21dsa1d2311bbbbs}) can be determined by
\begin{equation}
\begin{split}\label{bulkoutageprob}
&P_{out}^{bulk}(s)=({F}_I(s))^M=\frac{S_0(K,s,q)^{M(Kq+1)}}{(S_1(K,s,q)-S_0(K,s,q))^{MKq}}\\
&\times f_\gamma\left(\frac{S_1(K,s,q)-S_0(K,s,q)}{S_0(K,s,q)}\right)^{MK}+O\left(\frac{1}{\bar{\gamma}^{MK(q+1)+2}}\right).
\end{split}
\end{equation}
Furthermore, we can derive the asymptotic outage probability at high SNR by power series and obtain
\begin{equation}\label{asympoutagebulk}
P_{out}^{bulk}(s) \thicksim \tilde P_{out}^{bulk}(s)=\left({S_0(K,s,q)g_0^K}\right)^M.
\end{equation}
Note, because of the condition $g_1(\bar{\gamma})=o(g_0(\bar{\gamma}))$ when $\bar{\gamma}\rightarrow \infty$, the diversity order can be derived from (\ref{asympoutagebulk}) by $G_d^{bulk}=-\lim_{\bar{\gamma}\rightarrow\infty}\frac{\log P_{out}^{bulk}(s)}{\log \bar{\gamma}}=MK(q+1)$.

\subsection{Per-subcarrier selection}
According to (\ref{defoutage}) and (\ref{ptsc}), the \textit{a posteriori} outage probability when per-subcarrier selection is employed can be defined by
\begin{equation}\label{sajhfkjh1jfdjhmamamji}
\begin{split}
&P_{out}(s)=\mathbb{P}\left\lbrace \sum_{k=1}^{K} \ln (1+\max_{m\in\mathcal{M}}\gamma(m,k))<s \right\rbrace.
\end{split}
\end{equation}
Denote $\Psi(m,k)=\max_{m\in\mathcal{M}}\gamma(m,k)$. We can obtain the CDF and PDF of $\Psi(m,k)$ as
\begin{equation}
F_\Psi(s)=\left(F_\gamma(s)\right)^M\Leftrightarrow f_\Psi(s)=M \left(F_\gamma(s)\right)^{M-1}f_\gamma(s).
\end{equation}
Following the assumption given in (\ref{56465122211}), $f_\Psi(s)$ can be further expressed by
\begin{equation}
\begin{split}
&f_\Psi(s)=\frac{Mg_0^M}{(q+1)^{M-1}}s^{M(q+1)-1}\\
&~~+M\left[\frac{g_0^{M-1}g_1}{(q+1)^{M-1}}+\frac{(M-1)g_0^{M-1}g_1}{(q+1)^{M-2}(q+2)}\right]s^{M(q+1)}\\
&~~+O\left(s^{M(q+1)+1}\right).
\end{split}
\end{equation}
Therefore, if we denote
\begin{equation}
\begin{cases}
q'=M(q+1)-1\\
g_0'=\frac{Mg_0^M}{(q+1)^{M-1}}\\
g_1'=M\left[\frac{g_0^{M-1}g_1}{(q+1)^{M-1}}+\frac{(M-1)g_0^{M-1}g_1}{(q+1)^{M-2}(q+2)}\right]
\end{cases}
\end{equation}
$f_\Psi(s)$ can also be written as
\begin{equation}
f_\Psi(s)=s^{q'}(g_0'+g_1's+O(s^2)),
\end{equation}
which also aligns with the form given in (\ref{56465122211}). This result indicates that we can similarly employ \textit{Proposition 1} derived in \cite{7312903} to analyze the outage performance of per-subcarrier relay selection over the parallel fading channel, and the only modification is to replace $f_\gamma(\cdot)$ with $f_\Psi(\cdot)$. Hence, the outage performance of per-subcarrier relay selection over the parallel fading channel as defined in (\ref{sajhfkjh1jfdjhmamamji}) can be calculated by
\begin{equation}
\begin{split}\label{psoutageprob}
&P_{out}^{ps}(s)= \frac{S_0(K,s,q')^{Kq'+1}}{(S_1(K,s,q')-S_0(K,s,q'))^{Kq'}}\\
&\times f_\Psi\left(\frac{S_1(K,s,q')-S_0(K,s,q')}{S_0(K,s,q')}\right)^K+O\left(\frac{1}{\bar{\gamma}^{K(q'+1)+2}}\right).
\end{split}
\end{equation}
Furthermore, we can derive the asymptotic outage probability at high SNR by power series and obtain
\begin{equation}\label{asympoutageps}
P_{out}^{ps}(s) \thicksim \tilde P_{out}^{ps}(s)={S_0(K,s,q')g_0'^K}.
\end{equation}
Similarly to the case for bulk selection, we can also obtain the diversity of per-subcarrier selection from (\ref{asympoutageps}) to be $G_d^{ps}=MK(q+1)$ as expected.

\section{Applications}\label{app}
\subsection{DF relay networks}
According to (\ref{pdfchannelsingle}) and (\ref{snrdecodeandforward}), we can derive the PDF of the end-to-end SNR $\gamma_{DF}(m,k)$ for DF relay networks by
\begin{equation}\label{dsa54d51412xz}
f_{\gamma}^{DF}(s)=\frac{1}{\bar{\gamma}}\left(\frac{1}{\mu_1}+\frac{1}{\mu_2}\right)e^{-\frac{s}{\bar{\gamma}}\left(\frac{1}{\mu_1}+\frac{1}{\mu_2}\right)},
\end{equation}
which can be expanded by power series as $\bar{\gamma}\rightarrow \infty$ to the standard expanded form given in (\ref{56465122211}), and thus can be substituted into (\ref{bulkoutageprob}) and (\ref{psoutageprob}) to yield the approximated outage probability with bulk and per-subcarrier selections over the parallel fading channel.

\subsection{FG AF and VG AF relay networks}
By (\ref{pdfchannelsingle}) and (\ref{snrfixedgain}), we can similarly derive the PDF of the end-to-end SNR $\gamma_{FG}(m,k)$ for FG AF relay networks by \cite{7504171}
\begin{equation}\label{5465d4sa654da111111111}
\begin{split}
&f_{\gamma}^{FG}(s)=e^{-\frac{s}{\bar{\gamma}\mu_1}}\left[\frac{2(1+\bar{\gamma}\mu_1)}{\bar{\gamma}^2\mu_1\mu_2}K_0\left(2\sqrt{\frac{s(1+\bar{\gamma}\mu_1)}{\bar{\gamma}^2\mu_1\mu_2}}\right)\right.\\
&+\left.\frac{2}{\bar{\gamma}\mu_1}\sqrt{\frac{s(1+\bar{\gamma}\mu_1)}{\bar{\gamma}^2\mu_1\mu_2}}K_1\left(2\sqrt{\frac{s(1+\bar{\gamma}\mu_1)}{\bar{\gamma}^2\mu_1\mu_2}}\right)\right],
\end{split}
\end{equation}
where $K_x(\cdot)$ is the $x$th order modified Bessel function of the second kind.

Similarly, by (\ref{pdfchannelsingle}) and (\ref{snrvariablegain}), the PDF of the end-to-end SNR $\gamma_{VG}(m,k)$ for VG AF relay networks is given by \cite{7504171}
\begin{equation}\label{dsa09891}
\begin{split}
&f_{\gamma}^{VG}(s)=e^{-\frac{s}{\bar{\gamma}}\left(\frac{1}{\mu_1}+\frac{1}{\mu_2}\right)}\left[\frac{2(1+2s\mu_2)}{\bar{\gamma}^2\mu_1\mu_2}K_0\left(2\sqrt{\frac{s\left(1+s\mu_2\right)}{\bar{\gamma}^2\mu_1\mu_2}}\right)\right.\\
&\left.+\frac{2}{\bar{\gamma}}\left(\frac{1}{\mu_1}+\frac{1}{\mu_2}\right)
\sqrt{\frac{s\left(1+s\mu_2\right)}{\bar{\gamma}^2\mu_1\mu_2}}K_1\left(2\sqrt{\frac{s\left(1+s\mu_2\right)}{\bar{\gamma}^2\mu_1\mu_2}}\right)\right]
\end{split}.
\end{equation}

Although (\ref{5465d4sa654da111111111}) and (\ref{dsa09891}) cannot be expanded by power series to the form as given in (\ref{bulkoutageprob}), these two PDFs can still be substituted into (\ref{bulkoutageprob}) and (\ref{psoutageprob}) to yield the approximated outage probability with a divergent error at low SNR (see \textit{Corollary 3} in \cite{7312903}).

\section{Numerical Results}\label{nr}

\begin{figure}[!t]
\centering
\includegraphics[width=3.2in]{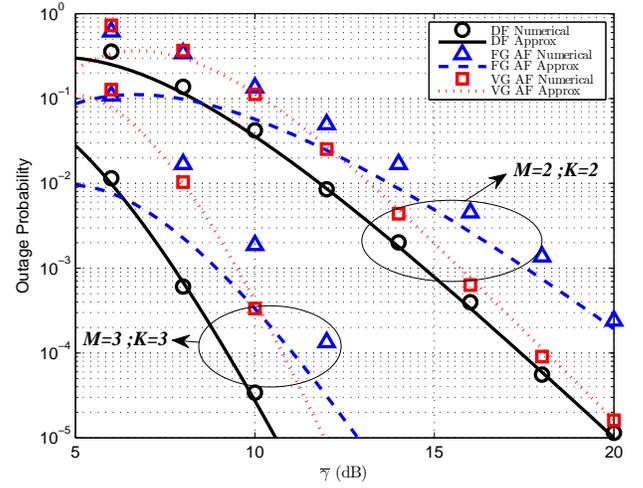}
\caption{Bulk selection case: numerical results and analytical approximations with different system configurations of $M$ and $K$.}
\label{bulksim}
\end{figure}

\begin{figure}[!t]
\centering
\includegraphics[width=3.2in]{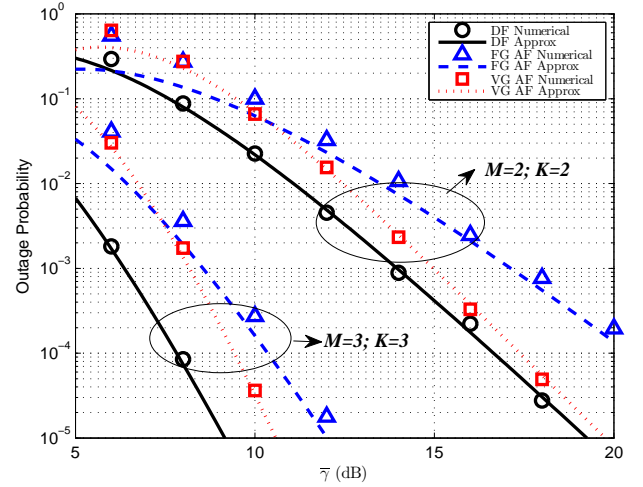}
\caption{Per-subcarrier selection case: numerical results and analytical approximations with different system configurations of $M$ and $K$.}
\label{pssim}
\end{figure}

To verify our analysis in Section \ref{opa} and Section \ref{app}, we carry out Monte Carlo simulations and present the numerical results in this section. To simplify simulations, without losing generality, we normalize the two-hop system by letting $\mu_1=\mu_2=1$ and $s=2$ (i.e. $\xi=1$ as the mutual information outage threshold). Then, we can plot the relation among $\bar{\gamma}$, analytical approximations for outage probability (as given in (\ref{bulkoutageprob}) and (\ref{psoutageprob})) and numerical outage probabilities for different combinations of $M$ and $K$. We present the simulation results for bulk selection and per-subcarrier selection in Fig. \ref{bulksim} and Fig. \ref{pssim}, respectively. From these two figures, it is clear that the proposed approximations for the outage probability with bulk and per-subcarrier selections over the parallel fading channel have been verified to be effective to approximate the outage performance of DF, FG AF and VG AF relay systems at high SNR. Note, although DF, FG AF and VG AF relay systems have the same diversity gain determined by $MK(q+1)$, the convergence rate of FG AF case to the asymptotic region is smaller due to different correction terms. Another minor point is that the similarity of outage performance between DF and VG AF relay systems in the high SNR region as valid over block fading channels \cite{7445895}, can still be found valid over the parallel fading channel. In addition, by comparing Fig. \ref{bulksim} and Fig. \ref{pssim}, we can find that the performance difference between bulk and per-subcarrier selections is not so significant as that in block fading channels \cite{7247508}. This is because the outage event over the parallel fading channel depends on the mutual information over all subcarriers, instead of each individual subcarrier.

\section{Conclusion}\label{c}
In this paper, we analyzed multicarrier relay selection for two-hop OFDM systems and obtained closed-form approximations for the outage probabilities when applying bulk and per-subcarrier relay selections. It has been numerically shown that the derived approximations for DF, FG AF and VG AF relay networks are valid and able to effectively approximate the exact outage performance at high SNR. Meanwhile, we also derived the generic asymptotic expressions for the outage probabilities at high SNR when applying bulk and per-subcarrier relay selections. All these results provide an insight into the
outage performance of multicarrier relay systems over parallel
fading channels.

\section*{Acknowledgment}
This work was supported by the SEN grant (EPSRC grant number EP/N002350/1) and the grant from China Scholarship Council  (No. 201508060323).

\bibliographystyle{IEEEtran}
\bibliography{bib}

\end{document}